**Microstructure and Elastic Constants of Transition Metal Dichalcogenide Monolayers from Friction and Shear Force Microscopy**

*Xiaomin Xu[†], Thorsten Schultz[†], Ziyu Qin, Nikolai Severin, Benedikt Haas, Sumin Shen, Jan N. Kirchhof, Andreas Opitz, Christoph T. Koch, Kirill Bolotin, Jürgen P. Rabe, Goki Eda and Norbert Koch\**

Dr. X. Xu, T. Schultz, Dr. N. Severin, Dr. B. Haas, Dr. A. Opitz, Prof. C. T. Koch, Prof. J. P. Rabe, Prof. N. Koch
Institut für Physik & IRIS Adlershof
Humboldt-Universität zu Berlin
12489 Berlin, Germany
E-mail: nkoch@physik.hu-berlin.de

Z. Qin, Prof. G. Eda
Department of Physics
National University of Singapore
Singapore117542

Z. Qin
State Key Laboratory of Materials Processing and Die Mould Technology
Huazhong University of Science and Technology (HUST)
Wuhan 430074, China

S. Shen
Department of Statistics
Virginia Polytechnic Institute and State University
Blacksburg, VA 24061, USA

J. N. Kirchhof, Prof. K. Bolotin
Department of Physics
Freie Universität Berlin
14195 Berlin, Germany

Prof. N. Koch
Helmholtz-Zentrum für Materialien und Energie GmbH
Bereich Solarenergieforschung
14109 Berlin, Germany

[†]X. Xu and T. Schultz contributed equally to this work.

**Abstract**

Optical and electrical properties of two-dimensional transition metal dichalcogenides (TMDCs) grown by chemical vapor deposition (CVD) are strongly determined by their microstructure. Consequently, the visualization of spatial structural variations is of paramount importance for future applications. Here we demonstrate how grain boundaries, crystal orientation, and strain fields can unambiguously be identified with combined lateral force microscopy (LFM) and transverse shear microscopy (TSM) for CVD-grown tungsten disulfide (WS$_2$) monolayers, on length scales that are relevant for optoelectronic applications. Further, angle-dependent TSM measurements enable us to acquire the fourth-order elastic constants of monolayer WS$_2$ experimentally. Our results facilitate high-throughput and nondestructive microstructure visualization of monolayer TMDCs, insights into their elastic properties, thus providing an accessible tool to support the development of advanced optoelectronic devices based on such two-dimensional semiconductors.

Transition metal dichalcogenides (TMDCs) are a class of two-dimensional (2D) semiconductors with in-plane covalent bonding and much weaker inter-plane van der Waals interaction.[1] An indirect-to-direct bandgap transition occurs when a bulk TMDC is thinned to a monolayer,[2,3] with sizable bandgaps of 1-2 eV.[4] The direct bandgap leads to strong light absorption and photoluminescence (PL), rendering TMDCs with their appealing optical and electronic properties[5,6] applicable in advanced nanoelectronics and optoelectronics.[7-9]

Large-area and industry-scale fabrication of TMDC monolayers seems within reach due to the rapid development of chemical vapor deposition (CVD).[10-14] Yet, spatial structural variations, e.g., grain boundaries (GBs) and strain fields, exist naturally within individual flakes[15-17] and polycrystalline films,[18] and significantly alter the local optoelectronic properties. Likewise, fundamental research requires detailed knowledge of sample heterogeneity. Therefore, identifying the location and nature of GBs and unraveling the mutual crystal lattice orientation are needed in a nondestructive and time-efficient manner.[19] Detection of GBs is so far mostly realized by scanning tunneling microscopy, transmission electron microscopy, PL and Raman mapping,[20-23] which have an inherently low analysis throughput. Sample treatments, including controlled preferential oxidation[24] and $H_2O$ vapor etching of GBs,[25] were reported, providing possibilities to detect GBs by optical microscopy. Furthermore, non-linear optical methods, such as second-harmonic generation[26] and third-harmonic generation,[27] have recently been demonstrated as powerful tools that allow a rapid visualization of GBs in molybdenum disulfide ($MoS_2$) monolayers. However, none of these techniques offers the feasibility of acquiring information on GBs and lattice orientation simultaneously and efficiently on a large-area scale, yet with nanoscopic detail.

Scanning force microscopy (SFM) is capable of investigating mechanical properties of surfaces,[28] which in turn provides microstructure information. Lateral force microscopy (LFM) and transverse shear microscopy (TSM) are two of the SFM contact modes, employing a perpendicular and parallel scanning direction relative to the cantilever axis, respectively, and they have been instrumental in revealing friction anisotropy in molecular crystalline films[29] and graphene.[30] However, despite its ease of use, the general applicability of LFM and TSM to sense transverse shear, to discern the elastic modulus tensor, and thus to reveal the crystallinity and crystal orientation, has been unrecognized for 2D TMDCs to date.[31]

In this work, we demonstrate a facile approach to identify GBs and visualize the relative lattice orientation of 2D TMDC monolayers. The combination of LFM and TSM images can reliably unravel the microstructure, distinguish between single-grain and multi-grain flakes, and discriminate strain fields from structural defects due to diagnostic contrast. TSM is demonstrated to be superior in identifying grain orientation with high contrast. In particular, we show that angle-dependent TSM measurements allow the determination of minimal grain misorientation and an estimation of the fourth-order elastic constants of monolayer TMDCs, exemplarily demonstrated for $WS_2$.

**Results and Discussion**

We focus on monolayer $WS_2$ grown by a salt precursor initiated CVD (described in the Methods section) on $SiO_2$/Si substrates, which yields crystalline triangular-shaped flakes with a typical edge length varying from tens of micrometers up to 350 μm in our case. As shown in Figure S1 of the Supporting Information (SI), the topographical SFM image (acquired with tapping mode) of the majority of the $WS_2$ triangular flakes exhibits step heights of ~ 0.8 nm, and the Raman spectrum shows $E^1_{2g}$ and $A_{1g}$ phonon modes at 356 and 417.5 cm$^{-1}$, respectively ($\lambda_{ex}$ = 532 nm). A single PL peak at ~ 1.95 eV with a full-width at half-maximum (FWHM) of ~ 50 meV corresponds to the only direct excitonic transition at the K point of the Brillouin zone. All the above features are evidence of monolayer $WS_2$ with high quality.[32,33]

*Single-grain flakes*

We first discuss an as-grown triangular flake showing no contrast in LFM (Figure 1a) and TSM (Figure 1b) images, and exhibiting homogeneous photoluminescence in the interior (Figure 1c). Small fluctuations toward the crystal edge in the PL intensity map are most likely due to a residual random distribution of defects and defect-bound excitons.[34] The equilateral triangle with 60° angles suggest a structure with the triangle edges parallel to a specific lattice orientation.[21] The selected area electron diffraction (SAED) patterns revealed only one set of diffraction spots (Figure S2a of the SI), evidencing that this grain is indeed single-crystalline with hexagonal structure.

Next, we attend to an as-grown flake that exhibits a diffuse, star-shaped contrast in LFM that emanates from the flake center (Figure 2a). A typical deflection detector contrast of 5~10 mV was frequently observed in LFM images of numerous as-grown triangular flakes, an example shown in Figure 2b. Notably, the TSM image acquired from the same flake exhibited homogeneity throughout (Figure 2c). A PL quenching by ~ 60%, accompanied by a slight red shift of the optical absorption maximum, was observed for the locations with the diffuse contrast in LFM (Figure 2d-e). The diffuse LFM contrast disappeared after sample transfer onto another $SiO_2$/Si substrate (Figure 2f). Consequently, instead of having a structural origin, the diffuse contrast is most likely due to strain fields caused in as-grown (i.e., not transferred) single-grain flakes by the mismatched thermal expansion coefficients of TMDC and the $SiO_2$/Si substrate and the rapid cooling process after growth.[35]

Note that the flakes discussed in Figure 1 and 2 were all optically uniform, and undistinguishable from reflection microscopy images (Figure S2b-c in the SI). The diffuse star-shaped contrast in LFM images is attributed solely to changes of the friction coefficient of regions under strain. Compared to LFM, TSM can contrast different grain orientations, originating from elastic constant anisotropy.[29] Due to the threefold symmetry of the crystal structure of TMDCs, a homogeneous TSM image does not necessarily imply a single-grain flake, while a combination of LFM and TSM allows distinguishing between single-grain and multi-grain flakes, which is discussed in the following section.

*Multi-grain flakes*

Figure 3 displays SFM topography, LFM and TSM images of two monolayer $WS_2$ flakes: one with non-straight edges (Figure 3a-c) and another having equilateral geometry with 60° angles (Figure 3d-f), respectively. Differences in the triangle edge shapes were assigned to different edge terminations, i.e., straight edges for metal-zigzag edges than for S-zigzag ones.[20] The flakes discussed here are again optically uniform (see, e.g., reflection microscopy image in Figure 3a) with homogeneous height characteristic in the SFM topography (inset of Figure 3a). Opposed to the diffuse star-shaped LFM contrast attributed to strain (see above), in the LFM images of the two triangular flakes shown in Figure 3b, 3d and Figure S3a in the SI, star-shaped sharp lines pointing from each apex towards a common center appear, exhibiting a deflection signal contrast of about 30 mV (Figure S3b of the SI) as opposed to 5-10 mV. These sharp lines persist after transfer of the flake from the as-grown substrate to another clean $SiO_2$/Si substrate (Figure 3e), during which any tensile strain introduced from rapid cooling supposedly is released (also *c.f.* Figure 2).[22,35] To better understand the origin of the unique star-shaped line contrast in LFM images, we performed PL mapping of the flakes. Narrow regions extending radially from the center to each apex of the flakes appear with a reduction of PL intensity by ~50% (shown in Figure 4), and these regions coincide with the lines in LFM images. Furthermore, PL peak positions and FWHM maps (Figure S4a of the SI) do not show obvious contrast, suggesting that the flake is indeed not strained.[36] It has been documented through TEM studies that GBs in TMDCs are as narrow as a few lattice constants;[20,21] yet they are visible as prominent lines with diminished (or enhanced) PL intensity compared to pristine crystal regions.[37,38] Intra-flake GBs exhibiting a star-shape geometry were investigated for CVD-grown $MoS_2$ with nano-Auger spectroscopy,[37] and for $WS_2$ with conductive atomic force microscopy[39], respectively, showing spatial variation in the defect density. Our observations and the above considerations let us conclude that the sharp line contrast inside the triangular flakes indeed corresponds to inherent structural defects, i.e., GBs. In principle, GBs, namely line defects, have atom vacancies or atoms shared by adjacent grains, and thus exhibit an abrupt change in the local lattice parameters, which in turn influences the frictional coefficient [40] and reflect on the contrast in the cantilever deflection signal. The deflection signal is not necessarily "higher" at the GB region compared to that of the pristine crystal region. The cantilever response to the change in frictional coefficient is related to the scan direction (Figure S3b of the SI). As stated in the Methods section, the LFM signals presented here were obtained as the difference in lateral bending between scans in forward (trace) and backward (retrace) direction, to remove features originating from topography differences.

It is worth noting that for multi-grain flakes with intra-flake GBs we commonly observe a particle-like nucleation center in the triangle's center, which we did not find for single-grain flakes (Figure S4b-c of the SI). This suggests two different nucleation and growth mechanisms forming triangular geometries. According to the schematic illustration shown in Figure S4b, when a critical nucleus with the $WS_2$ unit cell is formed spontaneously, it grows to form the equilateral shape,[41,42] and thus it is single-crystalline with homogeneous optical and elastic characteristics. When the nucleation site is of nano-particulate nature, e.g., a partially sulfurized $WO_3$ core, clusters, or inevitable substrate contamination,[18] our synthesis conditions seem to favor a growth, where - starting from this center - dominantly three grains merge with 60° twin grain boundaries, yielding the angle of 120° between intersecting edges and giving rise to the star-shaped contrast[43] (illustrated in Figure S4c of the SI). Related observations have been made for multi-grain $MoS_2$ monolayer flakes,[20] containing several rotationally symmetric mirror twins and thus different overall flake shapes. The formation and type of GBs could be influenced during growth to accommodate for local nonstoichiometry of the sample.[44] Studies toward the atomic structure of the star-shaped intra-flake GBs and the detailed growth dynamics are certainly to be undertaken in the future.

Overall, from 20 triangular monolayer flakes grown in different batches (under the same condition) and with variable sizes, we observed two flakes exhibiting the sharp line contrast of GBs in LFM, six flakes exhibited no contrast in LFM, and the remaining 12 flakes showed the diffuse star-shaped contrast, which disappeared after sample transfer. This suggests a single-crystalline nature of 90% of our CVD-grown triangular flakes (schematic illustration in Figure S4d of the SI).

In addition to the star-shaped lines, from the LFM images in Figure 3 one can also distinguish the edge periphery from the flake interior region. The contrast of the edge periphery is likely due to disorder caused by atomic vacancies, as previously suggested from campanile nano-optical probing combined with nanoscale Auger elemental mapping.[37] In fact, atom voids are generally observable at the edge of $WS_2$ in scanning transmission electron microscopy (STEM) (Figure S4e of the SI).

One should also notice that triangular flakes exhibiting homogeneous TSM images can be either of single crystalline nature (Figure 1b and 2c) or of polycrystalline nature with 120° intraflake GBs (Figure 3c and 3f), due to the threefold rotation symmetry of the hexagonal lattice. To unambiguously identify the crystalline nature, a combined analysis with LFM and TSM is necessary. To this end, variable flake microstructure and the resulting LFM/TSM deflection contrast are summarized in Table 1, where the high symmetry GBs denote GBs between the intersected crystals with relative in-plane rotation of n·60° (n=0,1,2…).

More complex polycrystalline patchworks form when the growth starts from a common center while heading into random directions with inequivalent speeds, or when single-crystalline grains merge during growth.[20,21] Figure 5a displays reflection microscopy and LFM images of as-grown $WS_2$ monolayers forming various geometries, including "mountain",

"butterfly", "fishtail", etc. Again, the spatial variations in LFM images, i.e., bright and sharp line-shape contrasts, are consistent with the PL signal fluctuations (Figure S5 of the SI), indicating the location of GBs. Tilt angles of the GBs from neighboring grain lattices are illustrated in the LFM images in Figure 5a. We noticed that although the angle $\alpha$ (indicated in blue) between two adjacent crystal domain edges assumes random values, the GB angle $\beta$ (angle between the GB and one neighboring domain edge, indicated in white) tends to be $\alpha/2$ in most cases, presenting symmetric tilt boundaries that bisect neighboring grains when the adjacent grains grow with comparable rates.[45,46] This observation is consistent with a previous report which proposed that the connection of two domains and the formation of GBs could involve a process of reconstruction until reaching a mirror symmetry.[24]

We have shown that GBs that mediate the crystal orientation transitions between the grains are visible in LFM images; further, lattice orientation variation in TMDC monolayers can be identified by TSM, while it is not possible by conventional optical methods (Figure S6a), PL/Raman spectroscopy or SFM topographic imaging. For example, the multi-grain flake "butterfly" exhibits remarkable contrast in TSM image (Figure 5b), highlighting distinct crystallographic orientation of four individual grains. The contrast in TSM is clearly not of topographic origin, but relates to the grain orientation relative to the fast scan direction (scan vector). Elastic anisotropy generates shear stress on the probe tip transverse to its scan direction when the scan direction is not along a symmetrical axis. This distinguishes TSM from LFM by providing striking sensitivity in revealing crystallographic orientation. The high contrast is also observed for other TMDC material monolayers (see CVD-grown $MoS_2$ in Figure S6b of the SI), demonstrating a general applicability. Note that the absolute TSM deflection signal is generally smaller than that of LFM, but has the same magnitude as previously reported for an organic semiconductor thin film.[29]

*Elastic constants of WS$_2$*

When a sample with elastic anisotropy is rotated with respect to the scan direction, the TSM signal varies strongly due to the different shear forces acting on the tip. Therefore, angle-dependent TSM measurements, i.e., varying the scan vector with respect to the sample, can be used to determine the elastic constants of a material.[47] So far, elastic properties of TMDC monolayers were scarcely attended to, one example being modeling by density functional theory (DFT) for MoS$_2$ by Peng et al.;[48] experimental insight is presently lacking. Using angle-dependent TSM measurements, we are able to provide estimates for the fourth-order elastic constants of WS$_2$. The data displayed in Figure 6 were obtained by incrementally rotating the sample, while keeping the scan direction along the cantilever axis (indicated by white arrows in Figure 6), deflection set point, and scanning speed fixed. Figures 6a-c demonstrate that the TSM signal of the two grains (labeled 1 and 2 in the images) composing the flake changes with respect to the sample orientation relative to the scan direction. The angle-dependent TSM signals obtained for the two flakes are summarized in Figure 6d, revealing friction anisotropy with a periodicity of 60°, as expected for a surface with hexagonal lattice structure and threefold rotation symmetry. This further explains the homogeneity of the TSM images taken from the multi-grain triangular flakes having high symmetry (120°) GBs.

Based on theory of linear elasticity,[49] a relation between the TSM signal *T*, the angle between scan direction and an in-plane principal direction *θ*, and the fourth-order elastic constants $C_{ijkl}$, that is generally applicable to any material, was derived by Kalihari et al.[46] Considering the hexagonal symmetry of TMDCs (for derivation see Supporting Information), the TSM signal *T* can be described as follows:

$$T = G\{C_{1111}[\cos(3\theta + \alpha) \cdot \sin^3(3\theta + \alpha) - \cos^3(3\theta + \alpha) \cdot \sin(3\theta + \alpha)]$$
$$+ C_{3333}[2\cos^3(3\theta + \alpha) \cdot \sin(3\theta + \alpha) - 2\cos(3\theta + \alpha) \cdot \sin^3(3\theta + \alpha)]$$
$$+ C_{1122}[\cos^3(3\theta + \alpha) \cdot \sin(3\theta + \alpha) - \cos(3\theta + \alpha) \cdot \sin^3(3\theta + \alpha)]$$
$$+ C_{1133}[\cos^4(3\theta + \alpha) - \sin^4(3\theta + \alpha)]\} \quad (1)$$

Here $\alpha$ is the tilt angle between the two grains of 178° as measured from two adjacent edges, and the experiment-specific $G$ is a constant with dimension voltage/pressure, which takes into account the applied stress, cantilever-tip geometry, instrument sensitivity, and tip-sample contact area. Due to a lack of documented fourth-order elastic constants of WS$_2$ so far, we applied the elastic constants of MoS$_2$ obtained from DFT calculations [47] as an initial guess for the calculation of *T*. A subsequent fitting (solid lines in Figure 6d) exhibits good agreement with the measured TSM data for elastic constants of $C_{1111} \approx$ 15000 GPa, $C_{3333} \approx$ 7000 GPa, $C_{1122} \approx$ 500 GPa and $C_{1133} \approx$ −800 GPa. The constant *G* in this case is 3.4 mV/GPa and $\alpha$ is 178° ± n·60° (n = 0,1,2…), in agreement with the measured tilt angle. The dashed lines in Figure 6d show the calculated *T* for a decrease of $C_{1111}$ (red dashed line) or $C_{3333}$ (blue dashed line) by 40 %, while keeping the other elastic constants unchanged, demonstrating the sensitivity of the calculated TSM signal on the elastic constants. Despite the good agreement between the calculated *T* and the measured TSM signal, one must keep in mind that the obtained absolute values are not unambiguous due to the unknown constant *G*. However, one can still sustain the following relative relation between the in-plane elastic constants of WS$_2$: $C_{1111} \approx 2C_{3333}$, $C_{1111} \gg C_{1122}$ and $C_{1111} \gg C_{1133}$. These values and relations can be of relevance for subsequent theoretical calculations of elastic properties of WS$_2$.

Because the scan vector with respect to the crystallographic orientation and thus the TSM deflection signal show azimuthal periodicity, rotational TSM measurements allow quantifying the relative crystallographic orientation of multiple flakes from the relative shift of the peak position in the fitted curves, as the deflection signals were obtained with a common scan vector. This represents particular ease when the flakes possess irregular shape and the crystallographic orientation cannot be distinguished simply from the edge orientation. Quantitative characterization of grain orientation was so far mostly realized by dark-field transmission electron microscopy (DF-TEM) in conjunction with filtered electron diffraction,[20,50] which requires restricted sample transfer process and is applicable over relatively small areas only. Rotational TSM measurements constitutes an alternative tool in a nondestructive, rapid manner and over a relatively large scales.

**Conclusion**

In conclusion, our study presents a coherent picture of the microstructures in CVD-grown TMDC monolayers through a combination of lateral force and transverse shear microscopies. Single crystalline triangular flakes exhibit no contrast in LFM or TSM images. Strain in single crystal grains appears as diffuse contrast in LFM, which disappears upon releasing the strain. Multi-grain flakes, either with a triangular shape or with more complex geometries, show bright, sharp line-shape contrast in LFM images, indicating the location of grain boundaries. Relative crystallographic orientations in multi-grain flakes are clearly unraveled by TSM. Angle-dependent TSM measurements further enable us to find relations between the fourth-order elastic constants of $WS_2$ ($C_{1111} \approx 2C_{3333}$, $C_{1111} \gg C_{1122}$ and $C_{1111} \gg C_{1133}$). Our study exemplifies a rapid and non-destructive approach to probe the microstructure of TMDC monolayers, through which a controllable utilization of desired strain, grain boundaries and crystal relative orientations can be greatly facilitated.

**Experimental Section**

*Salt precursor initiated CVD method*: An aqueous solution of $Na_2WO_3$ (40 mg in 40 mL water) was deposited on $SiO_2/Si$ substrates, which were cleaned in an ultrasonic bath sequentially with acetone and isopropanol and blow-dried with nitrogen gas prior to the chemical vapor deposition process. Subsequently, the $SiO_2/Si$ wafers with the deposited thin film of the precursor $Na_2WO_3$ were transferred into a quartz tube reactor. 100 mg sulfur were introduced upstream at a low temperature region (~ 200 °C). The furnace was heated to 200 °C for 5 min under an argon flow of 50 sccm. Subsequently, the temperature was ramped to 850 °C at a rate of 20 °C/min and kept for 5 min for sulfurization at atmospheric pressure. The furnace was then switched off and the tube was allowed to cool rapidly under 500 sccm Ar flushing.

*Characterization of the as-grown monolayer samples*: The $WS_2$ monolayers were characterized by Raman and PL spectroscopies, performed in a confocal microscope (XploRA, Horiba Ltd.)-based Raman spectrometer using a 532 nm laser with a 100× objective. The laser spot size was ~ 1 μm and the spectra were acquired with a maximum laser power of 0.015 mW. The 520 $cm^{-1}$ phonon mode from the silicon substrate was used for calibration. For the mapping, both the excitation and collection optics remained fixed while the sample was moved in x and y directions. All measurements were performed in ambient conditions.

*Transfer of the as-grown $WS_2$ monolayers*: To release the mechanical strain inside the monolayer crystals, the as-grown $WS_2$ flakes on $SiO_2$ (300 nm thermally oxidized)/Si wafers were transferred to clean $SiO_2/Si$ wafers with the commonly used wet-transfer method. Firstly, the samples were coated with a thin film of poly (methyl methacrylate) (PMMA) (495 PMMA, Micro Chem), followed by etching of the growth substrate in a hot base (KOH) solution bath to lift off the $WS_2$/PMMA films. The $WS_2$/PMMA films were then fished up with the target substrate and baked for 10 min at 90 °C. The samples were finally immersed in dichloromethane for 30 min to completely remove PMMA.

*Lateral force microscopy and transverse shear microscopy*: LFM and TSM are two variants of contact mode SFM, in which the lateral bending of the cantilever is measured. The difference in both methods is the movement of the cantilever with respect to its long axis: in LFM the cantilever is moved perpendicular to its long axis, whereas in TSM the cantilever is moved parallel to its long axis. All measurements were conducted using a Bruker Icon SFM and CONTV-A cantilevers. The applied load was kept constant during the angle-dependent TSM measurements. The LFM and TSM signals were calculated as the difference in lateral bending between scans in forward (trace) and backward (retrace) direction, and thus features originating from height differences are removed.

**Supporting Information**
Supporting Information is available from the Wiley Online Library or from the author.

**Acknowledgements**
X. Xu and T. Schultz contributed equally to this work. This work was supported by the SFB951 (DFG). The authors thank Tim Opitz and Prof. Emil List-Kratochvil for technical support, and the LEMMA group from INAC institute of the CEA Grenoble for access to the TITAN Ultimate microscope and help in acquiring STEM images of the voids. X. Xu acknowledges the Alexander von Humboldt Foundation for funding, Dr. Shisheng Li, Dr. Patrick Amsalem, and Dr. Soohyung Park for helpful discussions. B. Haas and C. Koch acknowledge the DFG-funded core facility BerlinEM Network (grant nr. KO2911/13-1). K. Bolotin was supported by the ERC Staring grant No. 639739. G. Eda acknowledges the Singapore National Research Foundation for funding the research under medium-sized centre programme. G. Eda also acknowledges support from the Ministry of Education (MOE), Singapore, under AcRF Tier 2 (MOE2015-T2-2-123, MOE2017-T2-1-134) and AcRF Tier 1 (R-144-000-387-114).

Received: ((will be filled in by the editorial staff))
Revised: ((will be filled in by the editorial staff))
Published online: ((will be filled in by the editorial staff))

**References**

[1]  S. Manzeli, D. Ovchinnikov, D. Pasquier, O. V. Yazyev, A. Kis, *Nat. Rev. Mater.* **2017**, *2*, 17033.
[2]  K. F. Mak, C. Lee, J. Hone, J. Shan, T. F. Heinz, *Phys. Rev. Lett.* **2010**, *105*, 136805.
[3]  A. Splendiani, L. Sun, Y. Zhang, T. Li, J. Kim, C.-Y. Chim, G. Galli, F. Wang, *Nano Lett.* **2010**, *10*, 1271.
[4]  Q. H. Wang, K. Kalantar-Zadeh, A. Kis, J. N. Coleman, M. S. Strano, *Nat. Nanotechnol.* **2012**, *7*, 699.
[5]  G. Fiori, F. Bonaccorso, G. Iannaccone, T. Palacios, D. Neumaier, A. Seabaugh, S. K. Banerjee, L. Colombo, *Nat. Nanotechnol.* **2014**, *9*, 768.
[6]  F. H. L. Koppens, T. Mueller, P. Avouris, A. C. Ferrari, M. S. Vitiello, M. Polini, *Nat.*

*Nanotechnol.* **2014**, *9*, 780.
[7] O. Lopez-Sanchez, D. Lembke, M. Kayci, A. Radenovic, A. Kis, *Nat. Nanotechnol.* **2013**, *8*, 497.
[8] K. F. Mak, J. Shan, *Nat. Photonics* **2016**, *10*, 216.
[9] Z. Sun, A. Martinez, F. Wang, *Nat. Photonics* **2016**, *10*, 227.
[10] Y. Lee, X. Zhang, W. Zhang, M. Chang, C. Lin, K. Chang, Y. Yu, J. T. Wang, C. Chang, L. Li, T. Lin, *Adv. Mater.* **2018**, *24*, 2320.
[11] Y. Zhan, Z. Liu, S. Najmaei, P. M. Ajayan, J. Lou, *Small* **2016**, *8*, 966.
[12] Z. Cai, B. Liu, X. Zou, H.-M. Cheng, *Chem. Rev.* **2018**, DOI: 10.1021/acs.chemrev.7b00536.
[13] J. Zhou, J. Lin, X. Huang, Y. Zhou, Y. Chen, J. Xia, H. Wang, Y. Xie, H. Yu, J. Lei, D. Wu, F. Liu, Q. Fu, Q. Zeng, C.-H. Hsu, C. Yang, L. Lu, T. Yu, Z. Shen, H. Lin, B. I. Yakobson, Q. Liu, K. Suenaga, G. Liu, Z. Liu, *Nature* **2018**, *556*, 355.
[14] S. Li, Y.-C. Lin, W. Zhao, J. Wu, Z. Wang, Z. Hu, Y. Shen, D.-M. Tang, J. Wang, Q. Zhang, H. Zhu, L. Chu, W. Zhao, C. Liu, Z. Sun, T. Taniguchi, M. Osada, W. Chen, Q.-H. Xu, A. T. S. Wee, K. Suenaga, F. Ding, G. Eda, *Nat. Mater.* **2018**, *17*, 535.
[15] H. Liu, J. Lu, K. Ho, Z. Hu, Z. Dang, A. Carvalho, H. R. Tan, E. S. Tok, C. H. Sow, *Nano Lett.* **2016**, *16*, 5559.
[16] Y. Sheng, X. Wang, K. Fujisawa, S. Ying, A. L. Elias, Z. Lin, W. Xu, Y. Zhou, A. M. Korsunsky, H. Bhaskaran, M. Terrones, J. H. Warner, *ACS Appl. Mater. Interfaces* **2017**, *9*, 15005.
[17] M. R. Rosenberger, H.-J. Chuang, K. M. McCreary, C. H. Li, B. T. Jonker, *ACS Nano* **2018**, *12*, 1793.
[18] Q. Ji, Y. Zhang, Y. Zhang, Z. Liu, *Chem. Soc. Rev.* **2015**, *44*, 2587.
[19] J. Wang, H. Yu, X. Zhou, X. Liu, R. Zhang, Z. Lu, J. Zheng, L. Gu, K. Liu, D. Wang, L. Jiao, *Nat. Commun.* **2017**, *8*, 377.
[20] A. M. van der Zande, P. Y. Huang, D. A. Chenet, T. C. Berkelbach, Y. You, G.-H. Lee, T. F. Heinz, D. R. Reichman, D. A. Muller, J. C. Hone, *Nat. Mater.* **2013**, *12*, 554.
[21] S. Najmaei, Z. Liu, W. Zhou, X. Zou, G. Shi, S. Lei, B. I. Yakobson, J.-C. Idrobo, P. M. Ajayan, J. Lou, *Nat. Mater.* **2013**, *12*, 754.
[22] Z. Liu, M. Amani, S. Najmaei, Q. Xu, X. Zou, W. Zhou, T. Yu, C. Qiu, A. G. Birdwell, F. J. Crowne, R. Vajtai, B. I. Yakobson, Z. Xia, M. Dubey, P. M. Ajayan, J. Lou, *Nat. Commun.* **2014**, *5*, 5246.
[23] Y. Zhang, Y. Zhang, Q. Ji, J. Ju, H. Yuan, J. Shi, T. Gao, D. Ma, M. Liu, Y. Chen, X. Song, H. Y. Hwang, Y. Cui, Z. Liu, *ACS Nano* **2013**, *7*, 8963.
[24] Y. Rong, K. He, M. Pacios, A. W. Robertson, H. Bhaskaran, J. H. Warner, *ACS Nano* **2015**, *9*, 3695.
[25] J. Wang, X. Xu, R. Qiao, J. Liang, C. Liu, B. Zheng, L. Liu, P. Gao, Q. Jiao, D. Yu, Y. Zhao, K. Liu, *Nano Res.* **2018**, DOI: 10.1007/s12274-018-1991-2.
[26] X. Yin, Z. Ye, D. A. Chenet, Y. Ye, K. O'Brien, J. C. Hone, X. Zhang, *Science* **2014**, *344*, 488.
[27] L. Karvonen, A. Säynätjoki, M. J. Huttunen, A. Autere, B. Amirsolaimani, S. Li, R. A. Norwood, N. Peyghambarian, H. Lipsanen, G. Eda, K. Kieu, Z. Sun, *Nat. Commun.* **2017**, *8*, 15714.
[28] S. Kasas, G. Longo, G. Dietler, *J. Phys. D: Appl. Phys.* **2013**, *46*, 133001.
[29] V. Kalihari, E. B. Tadmor, G. Haugstad, C. D. Frisbie, *Adv. Mater.* **2008**, *20*, 4033.
[30] J. S. Choi, J.-S. Kim, I.-S. Byun, D. H. Lee, M. J. Lee, B. H. Park, C. Lee, D. Yoon, H. Cheong, K. H. Lee, Y.-W. Son, J. Y. Park, M. Salmeron, *Science* **2011**, *333*, 607.
[31] M. Li, J. Shi, L. Liu, P. Yu, N. Xi, Y. Wang, *Sci. Technol. Adv. Mater.* **2016**, *17*, 189.

[32] H. R. Gutiérrez, N. Perea-López, A. L. Elías, A. Berkdemir, B. Wang, R. Lv, F. López-Urías, V. H. Crespi, H. Terrones, M. Terrones, *Nano Lett.* **2013**, *13*, 3447.
[33] C. Cong, J. Shang, X. Wu, B. Cao, N. Peimyoo, C. Qiu, L. Sun, T. Yu, *Adv. Optical Mater.* **2014**, *2*, 131.
[34] V. Carozo, Y. Wang, K. Fujisawa, B. R. Carvalho, A. McCreary, S. Feng, Z. Lin, C. Zhou, N. Perea-López, A. L. Elías, B. Kabius, V. H. Crespi, M. Terrones, *Sci. Adv.* **2017**, *3*, e1602813.
[35] W. H. Chae, J. D. Cain, E. D. Hanson, A. A. Murthy, V. P. Dravid, *Appl. Phys. Lett.* **2017**, *111*, 143106.
[36] H. J. Conley, B. Wang, J. I. Ziegler, R. F. Haglund, S. T. Pantelides, K. I. Bolotin, *Nano Lett.* **2013**, *13*, 3626.
[37] W. Bao, N. J. Borys, C. Ko, J. Suh, W. Fan, A. Thron, Y. Zhang, A. Buyanin, J. Zhang, S. Cabrini, P. D. Ashby, A. Weber-Bargioni, S. Tongay, S. Aloni, D. F. Ogletree, J. Wu, M. B. Salmeron, P. J. Schuck, *Nat. Commun.* **2015**, *6*, 7993.
[38] A. Alharbi, D. Shahrjerdi, *Appl. Phys. Lett.* **2016**, *109*, 193502.
[39] M. R. Rosenberger, H.-J. Chuang, K. M. McCreary, C. H. Li, B. T. Jonker, *ACS Nano* **2018**, *12*, 1793.
[40] D. H. Buckley, *Surface Effects in Adhesion, Friction, Wear, and Lubrication*, Elsevier, **1981**.
[41] A. Govind Rajan, J. H. Warner, D. Blankschtein, M. S. Strano, *ACS Nano* **2016**, *10*, 4330.
[42] H. Li, Y. Li, A. Aljarb, Y. Shi, L.-J. Li, *Chem. Rev.* **2017**, DOI: 10.1021/acs.chemrev.7b00212.
[43] X. Liu, I. Balla, H. Bergeron, M. C. Hersam, *J. Phys. Chem. C* **2016**, *120*, 20798.
[44] H.-P. Komsa, A. V. Krasheninnikov, *Adv. Electron. Mater.* **2017**, *3*, 1600468.
[45] J. Cheng, T. Jiang, Q. Ji, Y. Zhang, Z. Li, Y. Shan, Y. Zhang, X. Gong, W. Liu, S. Wu, *Adv. Mater.* **2015**, *27*, 4069.
[46] S. Park, M. S. Kim, H. Kim, J. Lee, G. H. Han, J. Jung, J. Kim, *ACS Nano* **2015**, *9*, 11042.
[47] V. Kalihari, G. Haugstad, C. D. Frisbie, *Phys. Rev. Lett.* **2010**, *104*, 169905.
[48] Q. Peng, S. De, *Phys. Chem. Chem. Phys.* **2013**, *15*, 19427.
[49] P. L. Gould, *Introduction to linear elasticity*, Springer-Verlag, **1983**.
[50] C. Luo, C. Wang, X. Wu, J. Zhang, J. Chu, *Small* **2017**, *13*, 1604259.

**Table 1**. Microstructure of TMDC monolayer flakes and the corresponding deflection contrast in LFM and TSM images.

| microstructure | LFM | TSM |
| --- | --- | --- |
| single crystalline | no contrast | no contrast |
| polycrystalline with high symmetry GBs[a] | sharp line contrast | no contrast |
| polycrystalline with low symmetry GBs[b] | sharp line contrast | contrast |

[a] High symmetry GBs denote GBs between the intersected crystals with relative in-plane rotation of n·60° (n=0,1,2…);
[b] Low symmetry GBs denote GBs between the intersected crystals with other tilt angle.

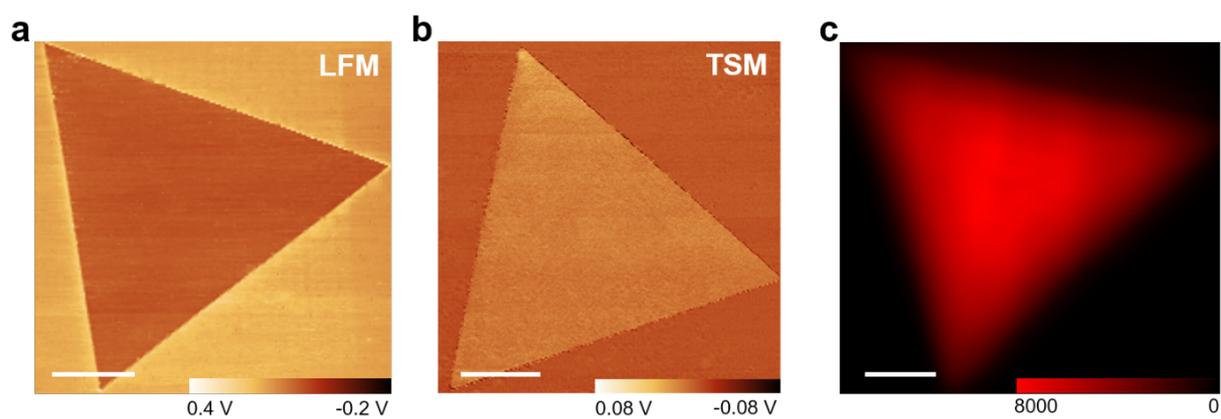

**Figure 1.** As-grown single-grain flake. a)-b) LFM and TSM images of a single-grain flake showing no contrast. c) PL intensity map obtained from the same flake with the color-scale in counts. Scale bar: 10 µm.

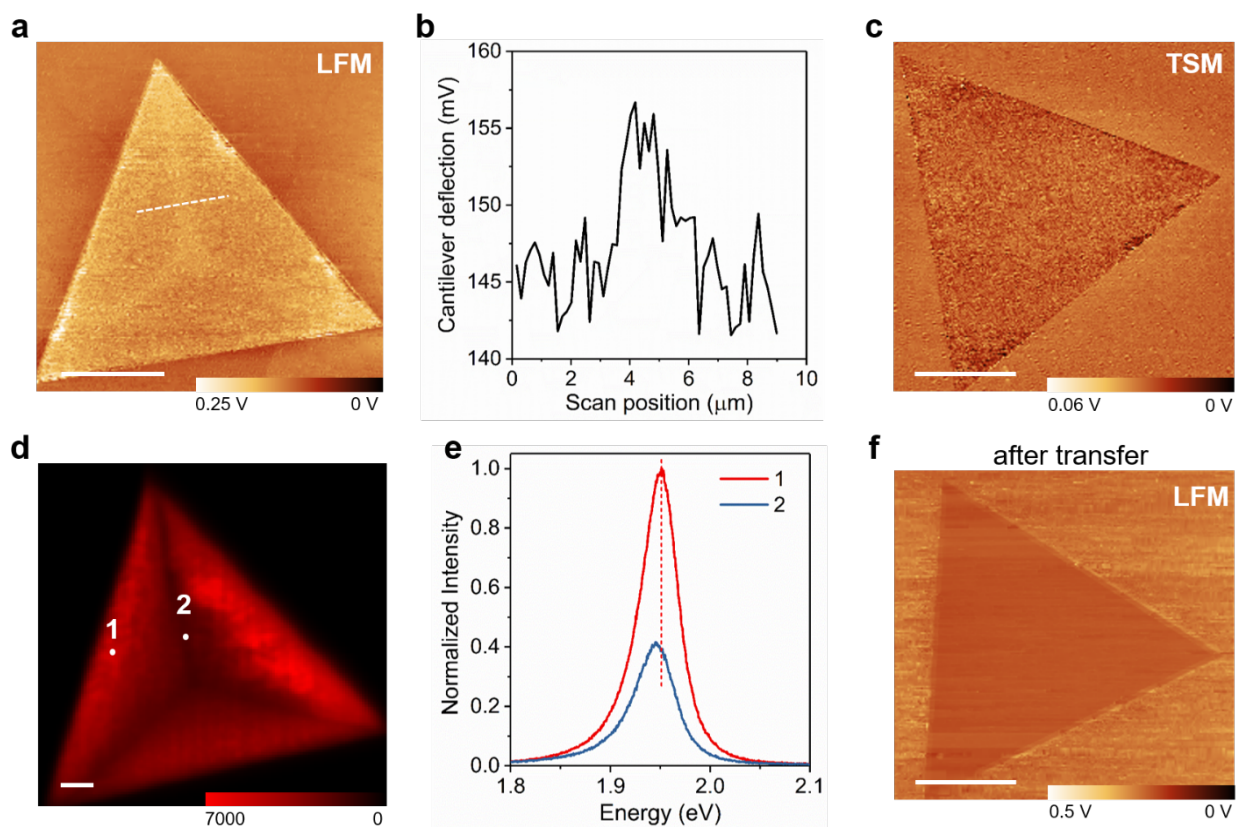

**Figure 2.** Single-grain flake under strain field. a) LFM image of an as-grown flake on SiO$_2$/Si substrate showing diffuse contrast characteristics of strain field. b) Line profile of the marked line in a), showing deflection signal of 5~10 mV difference in the location with diffuse contrast. c) TSM image of the same as-grown flake in a) showing no contrast. d) PL intensity map of a typical flake showing heterogeneity in photoluminescence as influenced by strain field (color-scale in counts). e) PL spectra of the spots marked in d), showing fluorescence quenching with red shift as an effect of strain. f) LFM image of the flake shown in a)-b) after transfer onto another SiO$_2$/Si substrate. Scale bar: 10 μm.

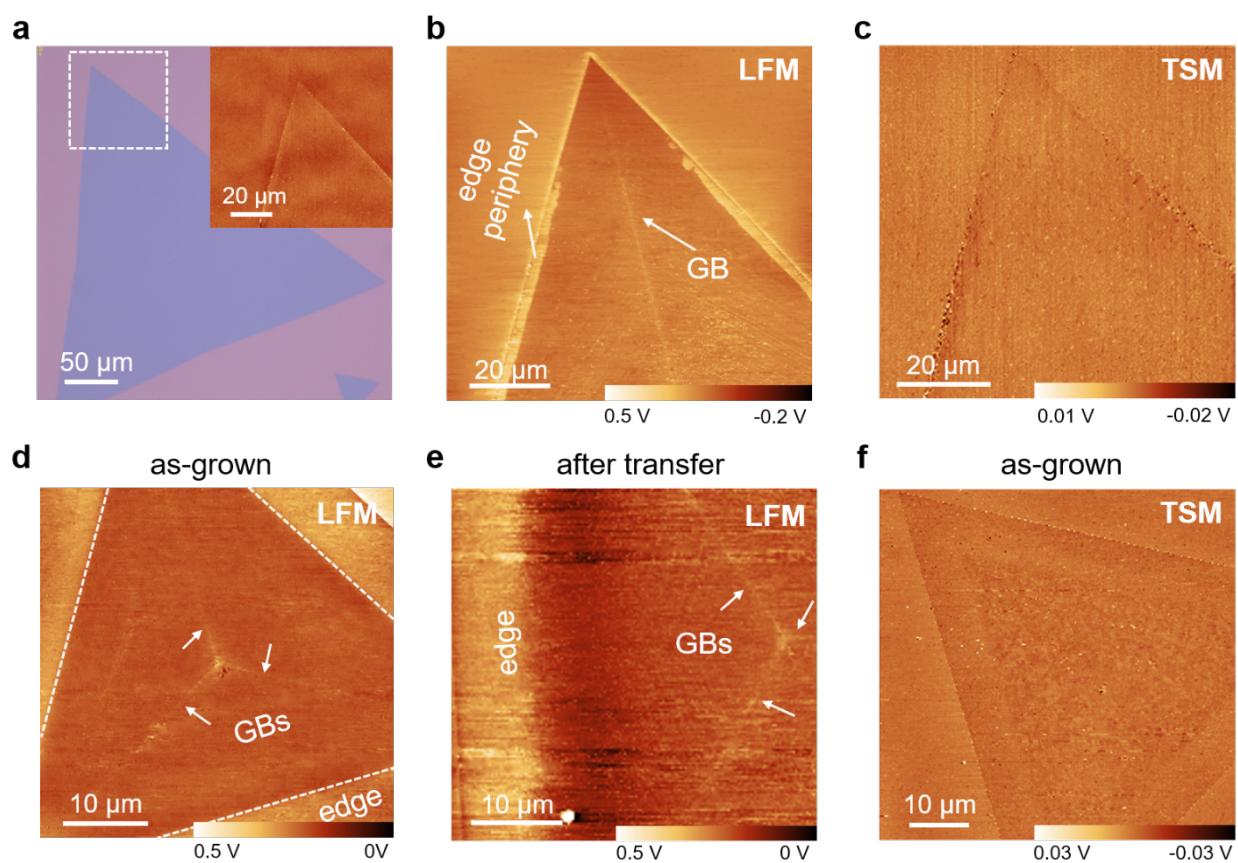

**Figure 3.** Multi-grain triangular flakes. a) Reflection microscopy image of an as-grown triangular-shaped WS$_2$ flake. Inset is the SFM topography image of one apex indicated by the white square frame, exhibiting homogeneous height with the thickness corresponding to a monolayer. b)-c) LFM and TSM images obtained from the same apex of the flake in a). d)-e) LFM images of an equilateral triangular flake before and after transfer. f) TSM image of the same flake in d).

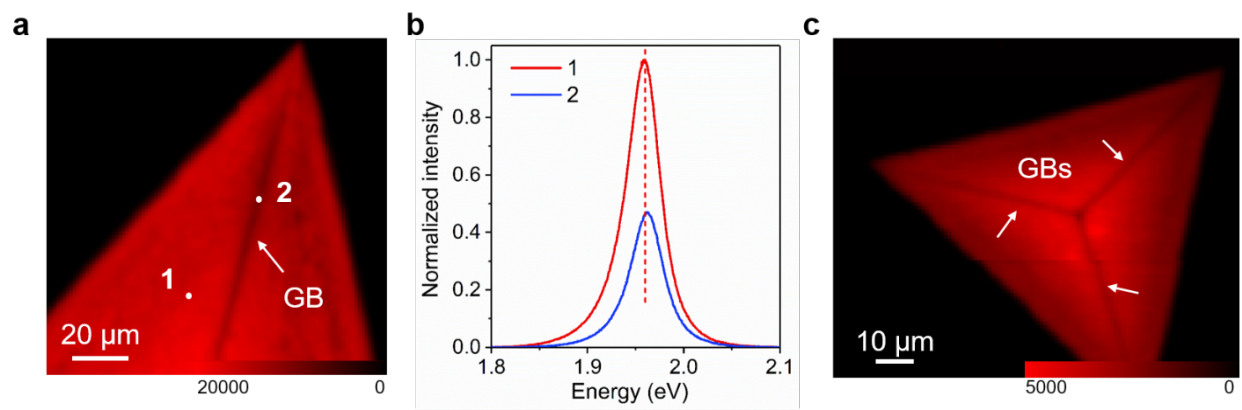

**Figure 4.** PL intensity maps of the multi-grain triangular flakes (color-scale in counts). a) PL intensity map obtained from the same apex as in Fig. 3a)-c). b) Point PL spectra of the marked spots in a) (1: flake pristine region; 2: location of GB). c) PL intensity map of the same flake shown in Fig. 3d)-f).

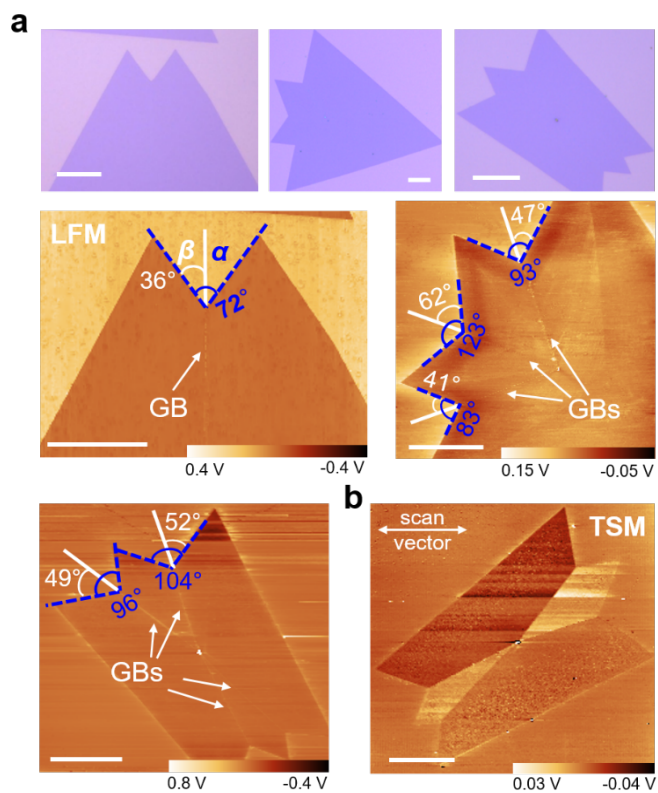

**Figure 5.** Multi-grain flakes. a) Reflection microscopy and LFM images of the multi-grain flakes forming the shape of "mountain", "butterfly" and "fishtail", respectively. The domain angle $α$ (angle between two neighboring edges) and GB angle $β$ (angle between the GB and the neighboring domain edge) are schematically illustrated in blue and white, respectively. b), TSM image of the "butterfly"-shape multi-grain flake. Scale bar: 20 μm.

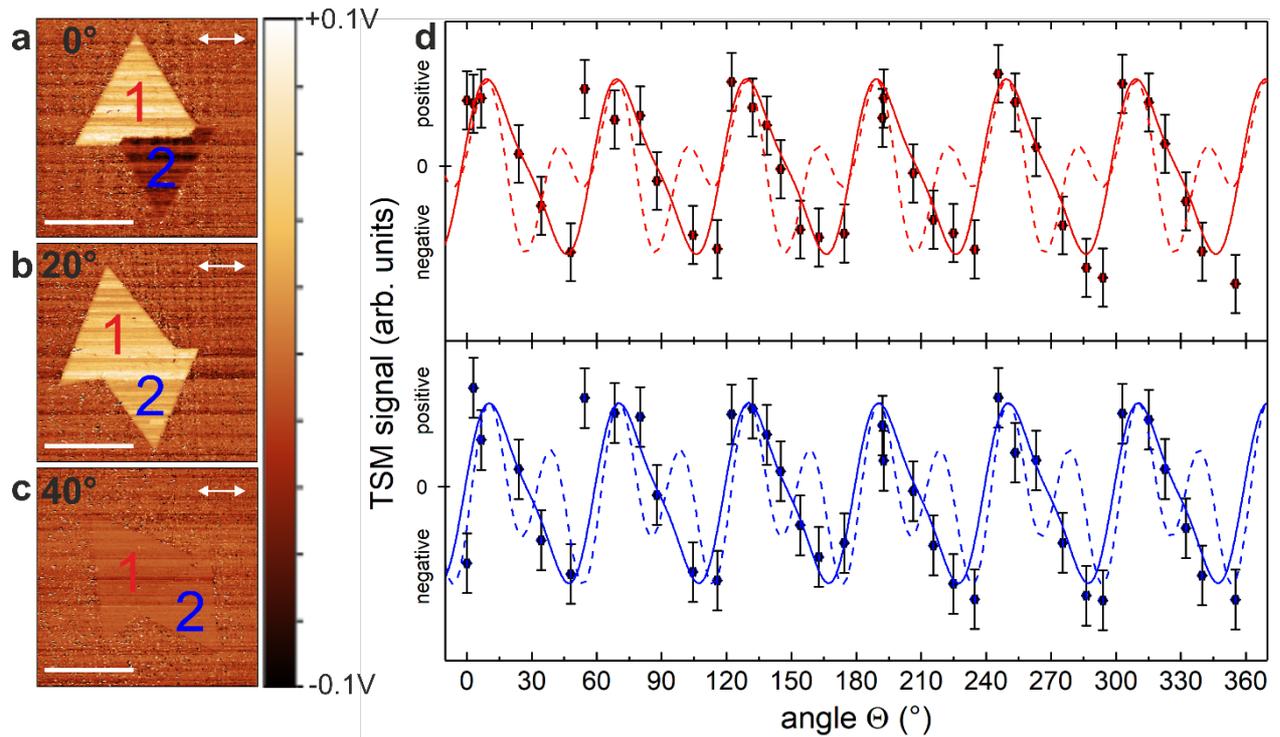

**Figure 6.** Angle-dependent TSM measurements on $WS_2$ flakes. Images a)-c) show exemplarily how the TSM signal changes drastically when the flake is rotated with respect to the measurement direction (indicated by white arrows). Scale bar: 20 µm. d) Summary of the TSM signal as a function of rotation angle $\theta$ for the two flakes indicated in a)-c). The solid lines show the simulated TSM signal for $C_{1111} \approx 15000$ GPa, $C_{3333} \approx 7000$ GPa, $C_{1122} \approx 500$ GPa and $C_{1133} \approx -800$ GPa. The dotted lines show the calculated TSM signals, if only $C_{1111}$ (red) or $C_{3333}$ (blue) are decreased by 40%, respectively, indicating the sensitivity of the TSM signal to variations in elastic constants.